\documentclass[journal=nalefd,manuscript=letter,layout=traditional]{achemso}
\usepackage[version=3]{mhchem} 
\usepackage{siunitx}

\usepackage{color}
\usepackage{easyReview}
\usepackage[numbers]{natbib}
\usepackage{bibunits}
\defaultbibliographystyle{mod_unsrtnat}  
\defaultbibliography{blinkingjunctions_bibliography.bib,references.bib,references_simulation.bib,references_konstantin_addition,biblioupdated.bib}

\author{Konstantin Malchow}
\email{konstantin.malchow@epfl.ch}
 \affiliation{Laboratoire Interdisciplinaire Carnot de Bourgogne CNRS UMR 6303, Université de Bourgogne, 21000 Dijon, France}
 \altaffiliation{Current adress: Laboratory of Quantum Nano Optics, EPFL Lausanne, 1015 Lausanne, Switzerland} 
 \alsoaffiliation{These authors contributed equally: Konstantin Malchow, Till Zellweger, Bojun Cheng}
 \author{Till Zellweger}
 \affiliation{Integrated Systems Laboratory, ETH Zurich, 8092 Zurich, Switzerland}
 \alsoaffiliation{These authors contributed equally: Konstantin Malchow, Till Zellweger, Bojun Cheng}
 \author{Bojun Cheng}
\affiliation{Microelectronics Thrust, Function Hub, Hong Kong University of Science \& Technology, Guangzhou, China}
\alsoaffiliation{These authors contributed equally: Konstantin Malchow, Till Zellweger, Bojun Cheng}
 \author{Aymeric Leray}
 \affiliation{Laboratoire Interdisciplinaire Carnot de Bourgogne CNRS UMR 6303, Université de Bourgogne, 21000 Dijon, France}
 \author{Juerg Leuthold}
\affiliation{Institute of Electromagnetic Fields (IEF), ETH Zurich, 8092 Zurich, Switzerland}

\author{Alexandre Bouhelier}
  \email{alexandre.bouhelier@u-bourgogne.fr}
\affiliation{Laboratoire Interdisciplinaire Carnot de Bourgogne CNRS UMR 6303, Université de Bourgogne, 21000 Dijon, France}

\title[]{Self-induced light emission in solid-state memristors replicates neuronal biophotons}

\begin{document}

\begin{abstract}
\noindent 

Key neuronal functions have been successfully replicated in various hardware systems. Noticeable examples are neuronal networks constructed from memristors, which are emulating complex electro-chemical biological dynamics such a neuron's efficacy and plasticity. Neurons are highly active cells, communicating with chemical and electrical stimuli, but also emit light. These so-called biophotons are suspected to be a complementary vehicle to transport information across the brain. Here, we show that a memristor also releases photons during its operation akin to the production of neuronal light. Critical attributes of biophotons such as self-generation, stochasticity, spectral coverage, sparsity and correlation with the neuron's electrical activity are replicated by our solid-state approach. Importantly, our time-resolved analysis of the correlated current transport and photon activity shows that emission takes place within a nanometer sized active area and relies on electrically induced single to few active electroluminescent centers excited with moderate voltage ($<$ 3V). Our findings further extend the emulating capability of a memristor to encompass neuronal optical activity and allow to construct memristive atomic-scale devices capable of handling simultaneously electrons and photons as information carriers. 

\end{abstract}

\begin{bibunit}

Solid-state neurons are essential elements to perform complex computation in brain-inspired processors. In the quest of implementing artificial neuronal networks, metal-oxide memristors have been deployed with great success to reiterate the principal components of neuronal cells activity~\cite{Jang21,sung_bio-plausible_2023}, including its wide range of plasticity~\cite{Chua18}, diffusive dynamics~\cite{Yang-NM2017}, connectivity and training~\cite{Strukoy15}. Memristors are two-terminal nanoscale nonlinear devices characterized by a conductance which depends on the bias history~\cite{Williams08}. The bio-mimetic advantage of memristive metal-oxide component is encoded in the resistive switching process, which emulate the electrochemical action potentials via the diffusion of charged species in the oxide matrix \cite{indiveri_reram-based_2016,sakai_multiple_2022}.

Another form of signal transmission in the brain has been proposed to take place since the discovery that living cells, including neurons, emit light as a result of their metabolic activities~\cite{Kobayashi99,Cifra:14}.
Biophotons are spontaneously emitted by reactive oxygen species produced by the mitochondria~\cite{Cifra:14}, and they may play a crucial role in the intrinsic cellular chemistry~\cite{Persinger14}. Biophoton emission covers a broad wavelength range spanning the visible to near-infrared~\cite{Chance81,Cifra:14}.  The light intensity is typically occurring at very low values~\cite{Inaba97} with rates varying from $10^2$ to $10^4$ photon$\cdot$sec$^{-1}$cm$^{-2}$. Biophotons are sometimes referred as ultraweak emission in the literature~\cite{Kobayashi99,Cifra:14}. The exact biological role of this light emission is still not clear~\cite{Cifra13}, but there is growing evidence that biophotons emitted by cerebral matter may act as supplementary information carriers to communicate the states of activity between neurons,~\cite{Mitrofanis21} and be part of the brain's learning process~\cite{Simon22}. 

In the context of the rapidly evolving bio-inspired integrated information processing, the additional capability of a solid-state neuron to emit and detect photons as a marker of its activity could be used as a signal to modulate information propagating in an artificial network. This non-local long range form of communication may enable the implementation of bio-inspired sophisticated learning algorithms\cite{Simon22}. For instance, light emission from pre-neurons can be collected and guided to post-neurons in order to modify their firing properties via a bio-plausible three-factor rule describing the effect of neuromodulators in the brain \cite{kusmierz_learning_2017,sarwat_chalcogenide_2022,portner_analog_2021}. 
Further, several papers have successfully realized photoactive or photogated memristive behavior~\cite{mao_photonic_2019,wang_optogenetics-inspired_2023}, and even implemented an optical readout~\cite{emboras_nanoscale_2013}. However, these demonstrators often involve complex hereto-integration of active materials and optical access limiting thus the intrinsic footprint of the structure. Additionally, emitted photons,  when present, are typically not correlated with the electrical activity of the devices.\\ 
In this study, we demonstrate that electrically-driven metal-oxide memristive devices used to replicate the neuron’s leaky-integrate and fire model \cite{indiveri_reram-based_2016,sakai_multiple_2022},
can also self release photons bearing attributes similar to biophotons produced during cell's actuation. In particular, we show that a planar \ce{Au}/\ce{SiO2}/\ce{Au} memristive junction emits photons during resistive switching cycles via a stochastic creation and activation of electroluminescent defect centers. By integrating and adapting methodologies from electrical transport analysis and optical time-resolved fluorescent measurements, we quantitatively analyze and describe the emitted photon efficiency, spectral coverage, stochastic dynamics and diffusion characteristics. Importantly, we demonstrate that our memristors act a nanoscale electrical light source at the level of single-to-few active electroluminescent emitters. We thus extract important photon parameters characterizing light-emitting memristors embedding electronic and photonic modalities in a single nanoscale device~\cite{Chen:14,Tappertzhofen:19} and provide a detailed quantitative characterization and analysis of light emitted by memristors operating in the trap-assisted tunneling regime.

\section{Results}
Figure~\ref{fig:IV}(a) introduces the device geometry used in this work. The memristor consists of an electrically biased planar Au optical gap antenna~\cite{Buret2015,Hecht15}, fabricated on a glass coverslip and covered by a sputtered 60 nm-thick \ce{SiO2} layer (see Methods section). The design of the gap antenna has been optimized to enhance both light emission and extraction (see Supplementary section). Device operation is investigated by a dedicated optical microscope equipped with small signal opto-electronic recovery equipment and sensitive photodetectors. Details of the setup are provided in the Supplementary section. The emulating synaptic behavior of such memristive devices has been described in a wide range of publications and is generally well understood \cite{sakai_multiple_2022,Cheng2019,lubben_active_2019}. Nonetheless, we remind here some basic working principle that are required to understand the emission of photons and their dynamics introduced later in the discussion. 

\subsection{Electroforming phase} 
\begin{figure}
    \centering
    \includegraphics[width=\linewidth]{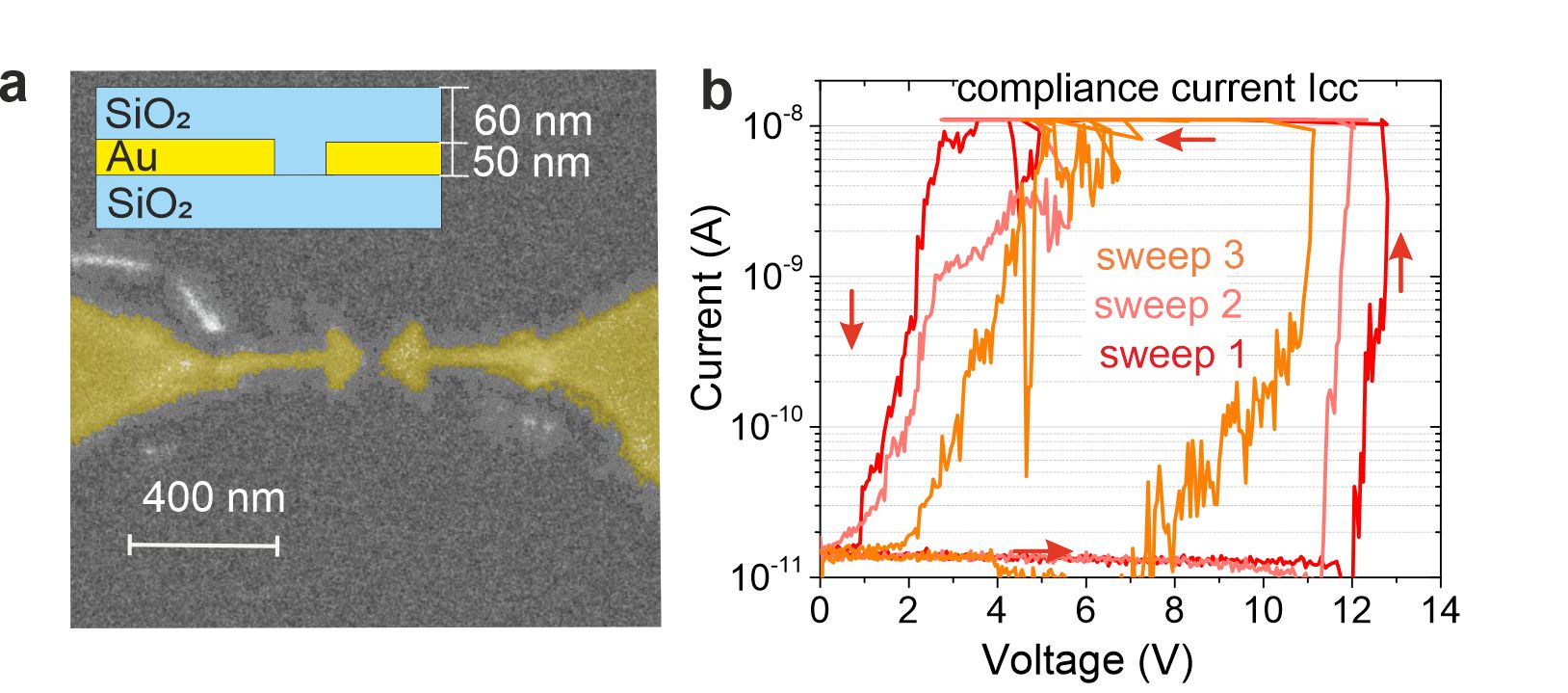}
    \caption{Device formation. (\textbf{a}) Scanning electron micrograph of the memristive device discussed here. The memristor is constituted of two 50 nm thick symmetric bow-tie shaped optical antennas separated by a gap of $\sim$35 nm. Microscopic Au electrodes are used for biasing the device. The entire structure is covered by thermally-sputtered 60 nm-thick \ce{SiO2}, as illustrated by the vertical layout. (\textbf{b}) is the current-voltage characteristics taken during 3 consecutive voltage sweeps used to electroform the device.  The compliance current $I_\textrm{cc}$ is constant.}
    \label{fig:IV}
\end{figure}

Pristine devices require an electroforming phase to undergo resisting switching. The purpose of this initial activation is to build conductive electron hopping paths between the two electrodes through defect creation and metal ion migration within the dielectric matrix~\cite{Yang2008,Wei12}. This defect injection mechanism is illustrated in Fig.~\ref{fig:IV}(b) showing a series of consecutive electrical voltage sweeps applied to electrically stress the junction. This process is described in greater details in the Supplementary section. Once a sufficient defect density is reached, a current flows in the dielectric medium by electrons hopping between localized states~\cite{Menzel19} and the device switches its conductance state akin to the firing of a neuron. These volatile conductive hopping paths are however thermodynamically unstable and the device relaxes to a high resistive state when the voltage $V_\textrm{{b}}$ is no longer applied. Once initiated, the resistive switching cycles occur recurrently, but the current characteristics evolve over time, switching at lower voltages to a high conductance state after repeated voltage sweeps, (see Supplementary information). After this electroforming phase, we operate the device with few-volts voltage pulses. By adjusting the pulse duration and duty cycle, the memristive junction can be repeatedly switched to a conductance range between \(10^{-8}-10^{-6}\)~S.
More in general, the value of the conductance can be further increased or decreased by the appropriate voltage drive, emulating potentation and depression behavior encountered in biological neural networks \cite{jimbo1999simultaneous}, see Supplementary Information.

\subsection{Photon emission from the memristor}

\begin{figure}
    \centering
    \includegraphics[width=\linewidth]{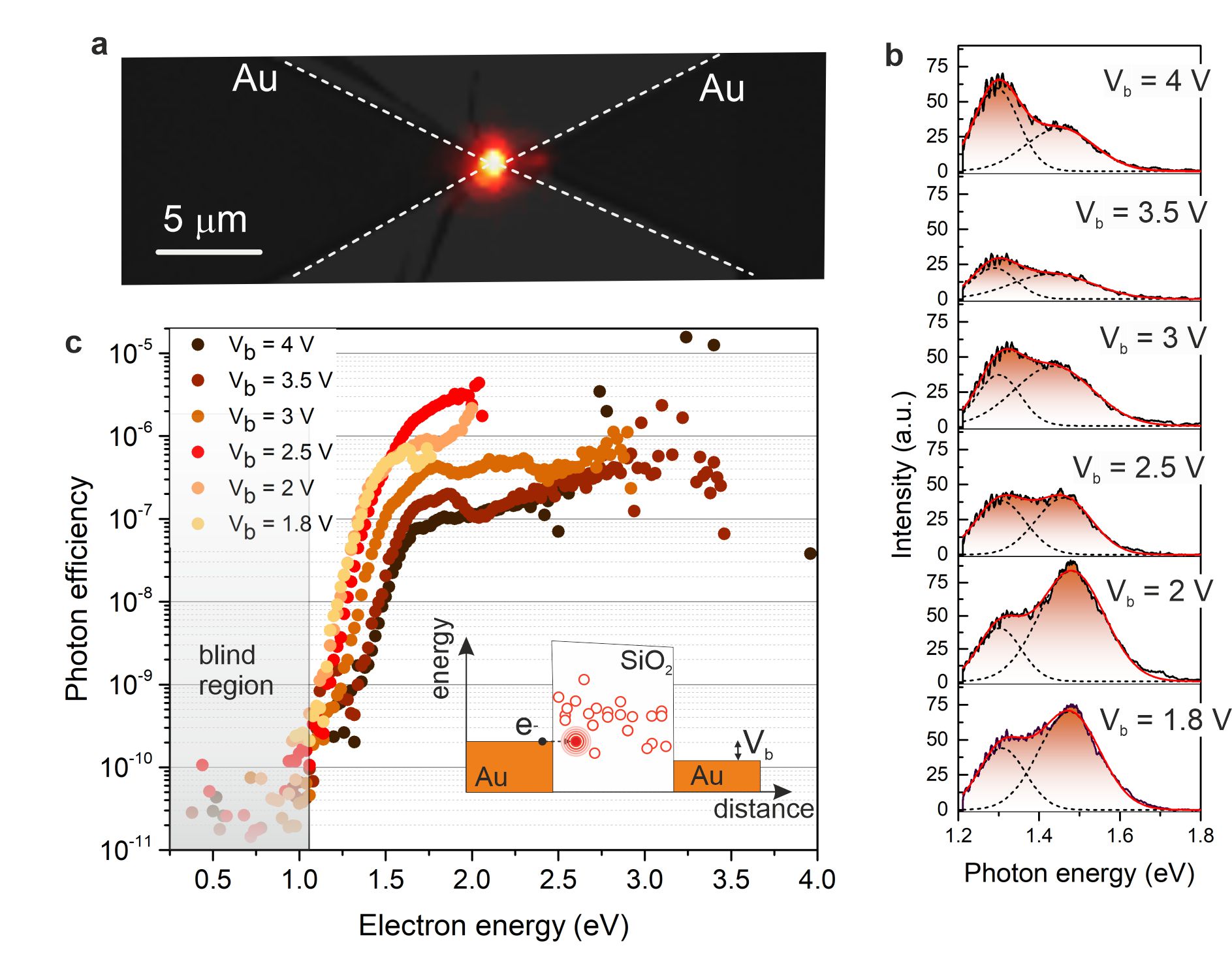}
    \caption{Light emission in the memristor. (\textbf{a})  Overlayed optical images of the device under a sweeping run. Photons are emitted from the switching region. The contour of the electrodes is made visible by the dotted lines.  (\textbf{b}) Series of emission spectra recorded for different pulse amplitudes. Each spectrum is integrated for 2 s. The dashed lines are Gaussian fits showing that the emitted spectra can be well described by only two resonances. (\textbf{c}) Photon efficiency, defined as the number of detected photons divided by the number of electrons, versus the kinetic energy of the electrons for a series of varied pulse amplitudes. The shaded area corresponds to the blind energy region of the APD. Inset: sketch of electron injection in recombining defect centers distributed in energy.}
    \label{fig:light}
    \end{figure}

Figure~\ref{fig:light}(a) shows an optical image of the photon emission from an activated memristor stressed by a voltage ramp as in Fig.~\ref{fig:IV}(c), overlayed with a bright field transmission image to observe the contour of the electrodes (dotted lines). Noticeably, light is emitted by the device within a diffraction-limited area overlapping the active region of the artificial neuron where resistive switching is taking place.  

Figure~\ref{fig:light}(b) shows the spectral distribution of the emitted photons measured for a series of voltage biases. Each spectrum has been recorded with an integration time of 2~s during a sequence of 10 ms pulses of constant amplitude. Clearly, the emission spectra can be reasonably described by two resonances at 1.29 $\pm$~8$\times10^{-3}$ eV and $1.46 \pm ~2\times10^{-2}$ eV with changing relative contributions between acquisition sequences. The persistent presence of both peaks shaping the emission indicate that two kinds of luminescent centers are responsible for light emission. A wide variety of intrinsic and extrinsic luminescent point defects can be simultaneously present in an amorphous \ce{SiO2} matrix, each emitting in separate energy windows~\cite{Salh2011}. Considering the spectral bands of the two peaks, it is likely that photons are emitted by a population of Si-nanoclusters of two different diameters~\cite{Salh2011}. Such Si-rich electroluminescent defects are known to be self-generated \emph{in-situ} during the switching process as a result of the formation and aggregation of oxygen vacancies~\cite{Tour:10,Zellweger:22,Cheng2022}. The presence of Au ions in the gap (see SI) may further influence the creation and electroluminescence of the Si-rich defects, we however observe very similar electroluminescence characteristics for devices featuring other metals and electrical conditions preventing metal ion migration into the \ce{SiO2} matrix \cite{Cheng2022}.
Coming back to the comparison with biophotons emitted by living cells, the near infrared spectral coverage shown here is analogous to the emission bands released by reactive oxygen species such as the high energy states of molecular oxygen~\cite{VanWijk:20}.   
In this picture, we can estimate the size of the Si nanocrystals to be in the order of 5~nm for the emission peak at 1.29 eV peak and approximately 8 nm for the peak at 1.46 eV using published size-dependent emission maximum~\cite{Liu:16}. We provide in the supplementary files spectra acquired from another device qualitatively showing similar spectra and dynamics. The relative contribution of the two populations depends on the peculiarity of the injection pathway involved when the memristor undergoes a sequence of resistive switching events. 

The emitted spectra coincide with the numerically calculated antenna resonance (see Supplementary section), probably contributing to an enhanced photon extraction efficiency. However, we also find that the efficiency strongly depends on the electrical potential difference and the kinetic energy given to the charge carriers. In Fig.~\ref{fig:light}(c), we plot the photon efficiency, defined as the number of photons counted in the spectral detection window divided by the number of electrons flowing in the device, versus the electron's kinetic energy. The carrier energy is deduced from the voltage effectively applied to the memristor $V_\textrm{{DUT}}$ (see supplementary information). The shaded area represents the blind spectral region of the photon detector, which has a cutoff energy at approximately 1.16 eV below which the detection efficiency vanishes. This cutoff energy coincides with the minimum bias required to detect a photon emitted through single-electron processes~\cite{Hamdad2022} such as electron injection in an emission defect center. Above this detection energy threshold, all curves feature a fast increase of the photon emission efficiency (3 to 4 orders of magnitude) up until approximately 1.8 eV. For higher voltages the efficiency reaches a plateau. Some curves show marked resonances even on its logarithmic representation. For instance, the data obtained for $V_\textrm{{b}}=3.5$~V feature a peaked efficiency at 1.8 eV. Such resonances are indicative of the energy levels of active emission centers bundled around specific energies within the bandgap of the \ce{SiO2} (see inset of Fig.~\ref{fig:light}(c)). Regarding the series as a whole, it clearly shows that further increasing the bias voltage improves the efficiency marginally: injected carriers likely loose their excess kinetic energy in the insulating layer by various other nonradiative decay mechanisms such as hopping between defects~\cite{Beasley:95}. This energy loss is also seen in Fig.~\ref{fig:light}(b), where a similar spectral coverage is detected regardless of the applied electron energies.
    
\subsection{Time dynamics and trap-induced fluctuations}

\begin{figure}
    \centering
    \includegraphics[width=\linewidth]{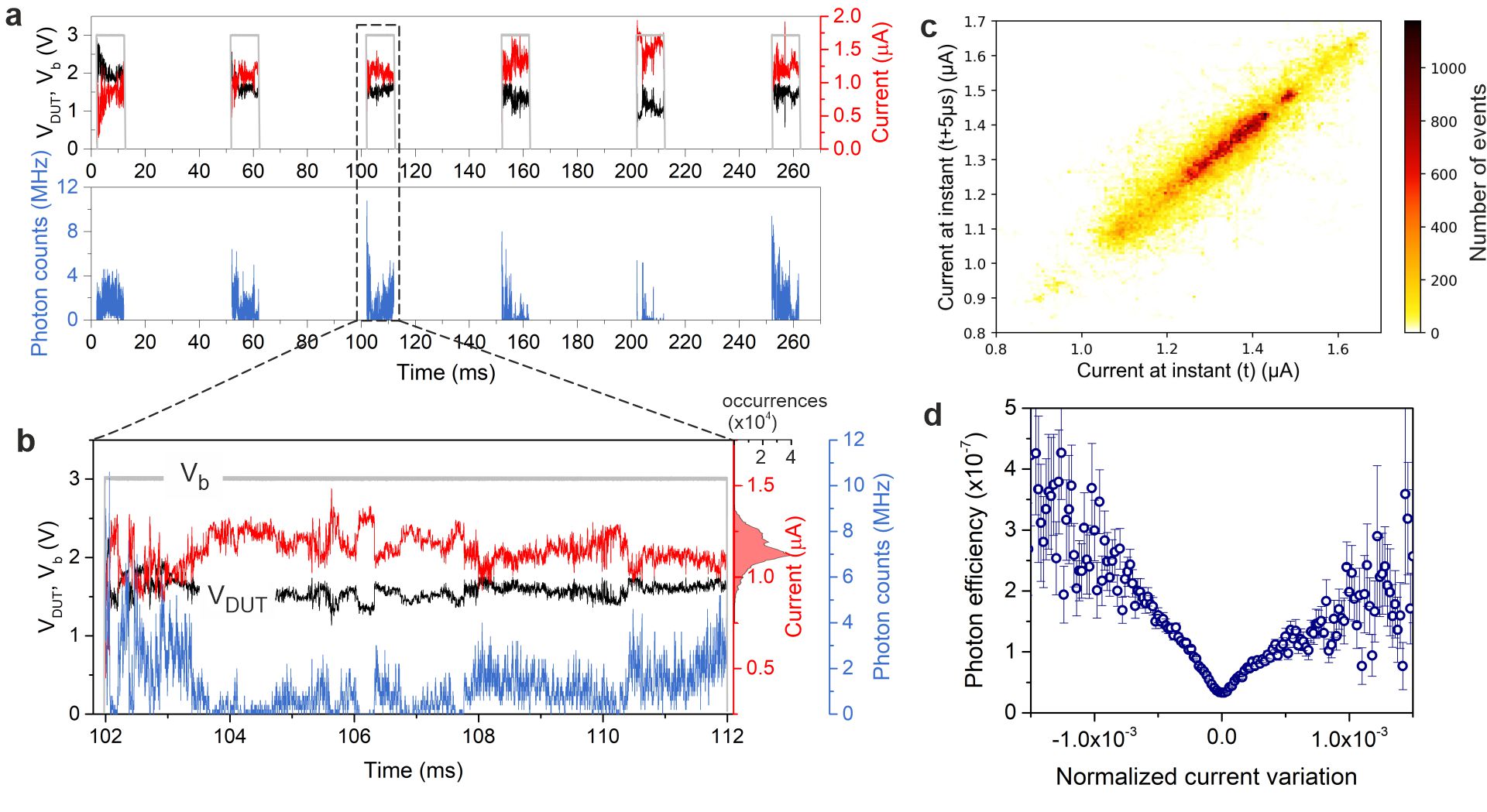}
        \caption{ Signals time dynamics. (\textbf{a}) Time traces showing the current, the photon counts and the voltage $V_\textrm{{DUT}}$ dropped at the gap for a series of $V_\textrm{b}=3$~V bias pulses (gray curve). (\textbf{b}) Details of the signals over a single pulse. The histogram of the current values is shown on the corresponding axis. (\textbf{c}) Current time lag plot extracted from the current evolution during a single pulse featuring a serially-dependent linear trend. (\textbf{d}) Photon efficiency versus relative current variation. Large current fluctuations generally give rise to a higher photon efficiency.
    }
    \label{fig:time}
    \end{figure}

Akin to biophoton flares signaling the transient physiological changes within the mitochondrion~\cite{VanWijk:20}, the (quasi)-static measurements shown in Figures~\ref{fig:IV} and \ref{fig:light} overlook that electrical and optical responses of the artificial neuron are subject to complex time dynamics. Both current and emission rates feature large instabilities with discontinuous spikes and sudden intensity jumps, which are intrinsically linked to trap-assisted tunneling and electron trap creation~\cite{Maturova2013}.
The local electric field produced by electrons captured in long lived shells are known to shift ground and excited state energies similar to a coulomb blockade\cite{Frantsuzov2008}. This reduces the number of available active traps inside the dielectric, and affects the devices' conductivity~\cite{Chen2001,Miranda2004,Wang2017}. As electrons are continuously captured and released from long-lived trap states, this dynamics is observed as a fluctuating current flowing in the artificial neuron, as shown in the red trace of Fig.~\ref{fig:time}(a). Here, a series of 10~ms voltage pulses with a period of 40~ms were applied to the device. The nominal bias amplitude was set to $V_\textrm{{b}}=3$~V (gray curve). The voltage dropped at the device $V_\textrm{{DUT}}$ is also shown in the time series (black curve). The blue curve is the photon activity produced by the memristor. In analogy with the biophotons signaling neuron's metabolism, a clear optical signal appears  when the device is active, that is when an electrical current flows through the junction and the memristor displays a relatively high conductance. To achieve current flow at these relatively low and short voltage pulses, the device was first conditioned by several voltage sweeps (see Supplementary Information).

Figure~\ref{fig:time}(b) shows with greater time resolution how these signals fluctuate within a single voltage pulse. The current shows stochastic noise and discrete jumps between multiple levels (see histogram of occurrences superposed on the current axis). These so-called gray states are indicative of the presence of several electron traps and defects~\cite{Lu:16}. The involvement of multiple traps is confirmed by the current pair occurrences \((I(t),I(t+5\mu s))\) featuring a linear distribution in the current time lag space~\cite{Ohata:17} shown in Fig.~\ref{fig:time}(c) for the same pulse. 
We observe a pulse-to-pulse variation of the number of states involved in the fluctuations of the current, with sometimes a clear signature of single traps dominating the current conduction and resulting into so-called telegraph noise (see Supplementary section).

The photons released by the memristor are apparently affected by the variation of the current. In Figs.~\ref{fig:time}(a) and (b), larger photon rates are generally observed concomitantly to sudden current changes. This flickering is most likely governed by trapping kinetics at the recombination centers~\cite{ElSayed:14}. This is better quantified in Fig.~\ref{fig:time}(d) where the photon efficiency is plotted versus the relative variation of the current. 
This directly translates to the number of recombining Si nanoclusters mobilized in the electrical transport. Hence, emitted photons are not reporting about the absolute activity of the artificial neuron (\emph{i.e.} the magnitude of the current flowing in the memristor), but are rather signaling the change in the transients of the conduction (action potential).

A clear confirmation of the critical role of localized traps in the transport and emission kinetics is the occurrence of a power law behavior in the distribution of so-called ON and OFF times~\cite{Huntley:06}. These times are defined when the current and photon count rate are above (ON) or below (OFF) a given threshold value. Such an analysis is usually conducted on the basis of photo-excited radiative systems, where the distribution of ON/OFF times reflects a time-dependent dispersion of trap energies. Here, we extend this methodology to electrically-excited memristors and apply it to both current and photon rate because traps are ruling both kinetics. Although the vicinity of the Au optical antenna may change the decay rate of the luminescent Si nanocrystals, the power law behavior was shown to be robust against the presence of metal contacts as well as the application of a bias~\cite{Bharadwaj:11}.  

\begin{figure}
    \centering
    \includegraphics[width=0.7\linewidth]{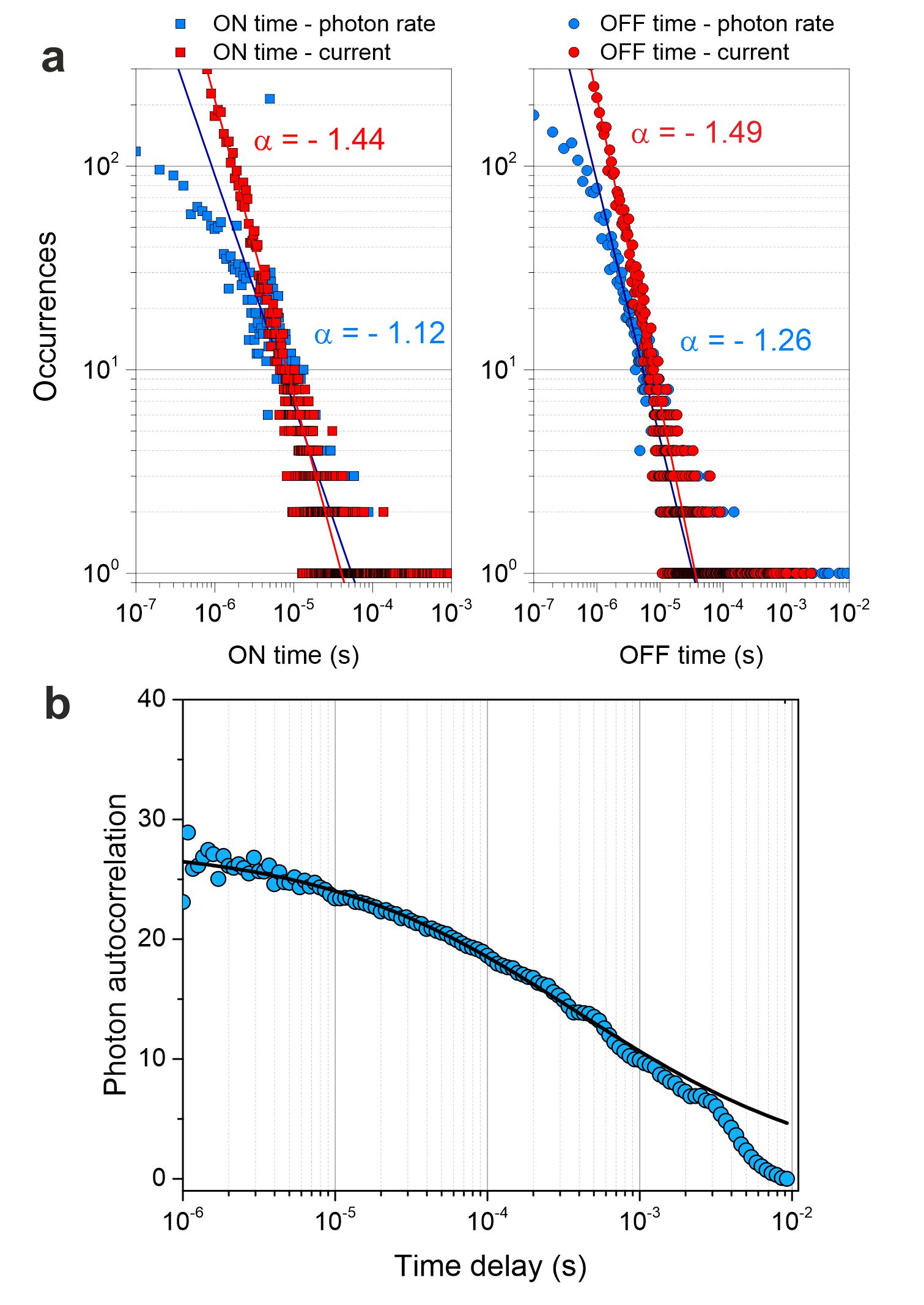}
    \caption{(\textbf{a}) Distribution of ON and OFF intervals of the current and the photon rate. The dashed lines are power law fits. The corresponding exponents $\alpha$ are indicated. (\textbf{b}) Second order autocorrelation $g^{(2)}(\tau)$ of the photon rate. The black line is a fit with an anomalous diffusion model.}
    \label{fig:powerlaw}
\end{figure}

Figure~\ref{fig:powerlaw} displays the number of occurrences of ON and OFF times deduced from of the time sequence of Fig.~\ref{fig:time}(a). For the current, we considered a threshold value equals to the mean current (here at 1.41 $\mu$A). For the photon rate, the threshold was set at 1.5 times the mean count rate mean (540 kHz) to limit distortion of the analysis~\cite{Cichos:14} and we excluded times below 5 $\mu$s from the analysis, which corresponds to the running average window applied to the data set. A power law of the form $t^{-\alpha}$ is clearly observed in these log-log representations of the two signals. The exponents $\alpha$ indicated in Fig.~\ref{fig:powerlaw} are extracted through standard fitting procedures (solid lines). These values, obtained from an electrically-powered device, are in line with blinking statistics observed in many photo-excited recombining centers, including Si nanocrystals~\cite{Huntley:06} and in systems featuring a distribution of states~\cite{Verberk2002}. We verified that the power law behavior is consistently retrieved and found a range of exponents comprised between -1.02 and -1.71.  We also checked that the statistics is robust against the threshold values~\cite{Bharadwaj:11,Cichos:14}.  Physically, the ON and OFF exponents are characteristic of the function ruling the trapping and escaping rates of the carriers~\cite{Searson:11}. Hence, observing a pulse-to-pulse variation of these rates during consecutive memristive switching redistributing the active centers in space and energy is not surprising because it reflects the atomic diffusion mechanism governing the electrical and optical responses of the artificial neuron.

Despite the apparent stochastic nature of the observed photon fluctuations, we are able to deduce important physical parameters such as the residence time and number of involved active centers by analysing the second-order auto-correlation function $g^{(2)}(\tau)$ of the fluctuating quantities. For the photon emission, the function is defined as: 
\begin{equation}
	g^{(2)}(\tau)=\frac{<I_\textrm{phot}(t)I_\textrm{phot}(t+\tau)>}{<I_\textrm{phot}(t)><I_\textrm{phot}(t+\tau)>}-1
	\label{eq:autocorrelation}
\end{equation}
where $I_\textrm{phot}$ is the photon rate at time $t$ and the brackets denotes the average over all times $t$. This function is well known in experiments aiming at determining the diffusion and reactivity of molecules and few emitters through time-resolved fluorescence correlation spectroscopy (FCS)~\cite{Enderlein:13}.   
With the 10 ns time resolution of our measurements the $g^{(2)}(\tau)$ can be inferred for sufficient short time delays.
Figure \ref{fig:powerlaw}(b) shows the auto-correlation trace estimated from timestamps extracted from all the voltage pulses of the cycling sequence applied to the memristor discussed in Fig.~\ref{fig:time}. Qualitatively, the steady decrease of $g^{(2)}(\tau)$ demonstrates that any photon correlation (\textit{i.e.} memory) will be lost after long enough time delays in this dynamically reconfiguring system. 

An important question immediately prompted by Fig. \ref{fig:powerlaw}(b) is which diffusion model can best account for the autocorrelation trace? The power-law statistics revealed in Fig. \ref{fig:powerlaw}(a) is a marker of an anomalous stochastic diffusion process taking place within the system~\cite{Metzler:14}. Amongst the different anomalous diffusion models, the continuous-time random walk provides a reasonable framework to discuss because it considers electron hopping through charge traps with random residence times~\cite{Metzler:14}. We restrict our analysis to a two-dimensional model because the out-of-plane electron diffusion is bounded to a good approximation by the thickness of the electrode (50~nm). With this assumption, the autocorrelation writes~\cite{Fradin:19}: 

\begin{equation}   
g_2(\tau)=\frac{1}{N}\frac{1}{1+\left( \frac{\tau}{\tau_{a}}\right)^{\beta}}
\label{eq:Anomalous diffusion model}
\end{equation}

where \(N\) is the average number of active emitters, \emph{i.e.}, the number of luminescent centers releasing photons at an instant $t$ averaged over the whole measurement. $N$ is given by the concentration $C$ of the fluctuating species in the detection volume $V$ ($N=CV$). \(\tau_a\) is the residence time and \(\beta\) is the anomalous diffusion exponent. \(\beta=1\) corresponds to normal Brownian diffusion. \(N\), \(\tau_a\), \(\beta\) are free parameters of the fit shown as a solid line in Fig.~\ref{fig:powerlaw}(b). Equation~\ref{eq:Anomalous diffusion model} agrees reasonably well with the experimental data using the parameters tabulated in Table~\ref{tab:autocorrelation fits}.

\begin{table}
    \centering
    \begin{tabular}{c|c|c|c}
         \(N\) &  \(\tau_a\)  & \(\beta\)  \\
        \hline
         0.0325(4) & 0.42(3) ms & 0.44(2) \\
    \end{tabular}
    \caption{The anomalous diffusion fit parameters deduced from Fig.~\ref{fig:powerlaw}(b). The numbers in brackets indicate the fit uncertainty in the last digits. }
    \label{tab:autocorrelation fits}
\end{table}

Although diffusion models have been developed for emitters diffusing in and out of a well defined detection volume in the context of Fluorescent Correlation Spectroscopy (FCS), the parameters inferred from the fit give a plausible picture of the transport as discussed below. The autocorrelation clearly confirms the anomalous subdiffusive character of the electron path with \(\beta\) smaller than 1. The number of emitters \(N\) needs to be corrected as the device is not excited continuously but through a series of voltage pulses. This is accounted for by simply normalizing  \(N\) with the pulse duty ratio \(\phi=0.2\), defined as the fraction of the pulse duration over the repetition period to determine the effective number of emitters \(N_\textrm{eff}\) during the activation of the device.
We find \(N_\textrm{eff}=N/\phi=0.16\). In average, less than one emitter is contributing to the emission of light from the device. This is consistent with the large photon rate fluctuations seen in the time traces and the inhomogeneous broadening of the time-integrated spectra (Fig.~\ref{fig:light}(b)). Using a diffusive circular area with a diameter given by the gap size, we estimate the concentration of emitting defects at $C=6.5\times10^{-3}$~nm$^{-2}$. In the Supplementary section, we verified the consistency of the model on an autocorrelation trace measured from a memristor stressed by a constant bias.

In the context of FCS molecular diffusion, \(\tau_a\) in Eq.~\ref{eq:Anomalous diffusion model} represents the average residence time of a fluctuating photo-excited emitter in the confocal measurement volume. However, in our device spatial diffusion of the electrons is restricted to inter-electrode spacing where the electric field is maximum. Consequently, the two-dimensional space available for trap creation and excitation is significantly smaller than the diffraction-limited detection area defined by the collection objective. We view \(\tau_a\) rather as the average active time during which a luminescence electron trap emits photons. Several factors are governing \(\tau_a\), including the formation, dissolution, migration and bleaching of the Si nanoclusters as well as their accessibility to electron charging under a dynamically changing conduction pathway. Revisiting the spectra of Fig.~\ref{fig:light}(b), the redistribution of the constitutive peak amplitudes suggests that, once formed, the luminescence centers remain stable longer than \(\tau_a\) but are switched on and off due to electron trapping and detrapping events modulating the conductive pathways.  \(\tau_{a}\) is  thus limited by the averaged electron trapping dynamics.

Knowing $N_\textrm{eff}$ and the detected photon rate $I_\textrm{phot}$, we can also infer the average emission rate per emitter $\gamma_\textrm{em}=I_\textrm{phot}/(N\eta_\textrm{coll})\simeq 4~\mathrm{MHz}$, where $\eta_\textrm{coll}$ is the detection efficiency of the apparatus (see Methods). This gives an average waiting time between photon events of $\tau_\textrm{em}=\gamma_\textrm{em}^{-1}=250$~ns. 
$\tau_\textrm{em}$ gives an upper limit of the excited state lifetimes of the Si nanoclusters responsible for the emission, which is a surprising value because the literature reports much longer decay kinetics extending well in the $\mu$s regime~\cite{Korgel:17}. However, the proximity of a plasmonic electrode design, the presence of metal ions in the dielectric layer, and carrier tunneling between the emitter and the dielectric matrix were all shown to decrease the radiative lifetime of Si nanocrystals~\cite{Cheng2022,Polman:05,Terukov:16}. Here, these effects are probably co-existing and, combined together, can reasonably explain the fast $\tau_\textrm{em}$. In average, the number of electrons $N_\textrm{excited}$ effectively trapped in the excited state of the recombining center must be small. $N_\textrm{excited}$ can be inferred from the emission rate $\gamma_\textrm{em}=N_\textrm{excited}\times\Gamma_\textrm{rad}$ where $\Gamma_\textrm{rad}=\eta/\tau$ is the radiative decay rate, $\eta$ is the quantum yield, and $\tau$ is the excited state lifetime. We do not have a direct measure of $\Gamma_\textrm{rad}$, but past photoluminescence studies evaluated the quantum yield at approximately 35\% for Si nanocrystals with size comprised between 4 to 8 nm~\cite{Korgel:17} and a lifetime measured at $\tau= 8$~ns in comparable electromagnetic environment~\cite{Cheng2022}. With these figures, we estimate $\Gamma_\textrm{rad}=43.75\times10^6$~s$^{-1}$ and $N_\textrm{excited}= \Gamma_\textrm{rad}/\gamma_\textrm{em}=0.09$ electron trapped in the excited state of the luminescent species when it is present.  


\section{Conclusion} 

We have shown here that metal-oxide memristors, which are commonly deployed to embody neuron's leaky-integrate and fire model can also emulate biophoton emission taking place in neuronal cells. Light emitted by the solid-state neuron shares similar characteristics to photons biologically released by neurons. Like chemiluminescent reactive oxygen species spontaneously produced by the cell's metabolic process, the active centers are self-generated during the actuation of the device as opposed to devices incorporating or associating semiconducting recombining layers~\cite{Malliaras:10}. The Si nanocrystals responsible for the light production in the artificial analogue emit in a spectral window overlapping that of biophotons. By adapting and developing time-resolved statistical analysis and methodology, we find that the optical activity of the memristor stems from a very low number of simultaneously active quantum defects, possibly single ones. Photon kinetics and correlation with the device's electrical response unambiguously revealed that photon production is associated with the fluctuation of the conductive state of the transmission channel (\textit{i.e.} the action potential of the neuron). However, any further quantitative comparison with biological neurons is difficult because \emph{in vitro} and \emph{in vivo} detection of biophotons at the cell level remains extremely challenging. 

Open questions and technical challenges remain to be addressed. For instance, it is not clear how to simultaneously implement in the artificial neuron of the type discussed here a photogated conduction that would enable an optical communication pathway between two units. Also, engineering such an optical transmission platform parallel to the electrical routing of a network of such devices demands more advanced fabrication strategies where flexible free-form additive manufacturing can play a decisive role~\cite{moughames_3d_2021}. Because our device operate (in average) with less than one quantum emitter, nontrivial statistics may become accessible to help constructing more advanced models for neural coding and processing. Finally, our approach may enable the implementation of an \textit{in-operando} optical monitoring of the computing dynamic activity of a network built from such memristors.

\section{Methods}
The devices are fabricated by electron beam lithography and subsequent ion milling on calibrated coverslips. First, a 1 nm thick Cr adhesion layer is thermally deposited in a vacuum chamber followed by electron-beam deposition of a 50 nm thick Au layer. A electron-sensitive resist (HSQ) is then spin coated on the surface and acts as a hard mask for the subsequent Ar milling of the pattern.  Finally, a 60 nm amorphous \ce{SiO2} cladding is deposited by radiofrequency sputtering and serves as the switching medium. 

The collection efficiency $\eta_\textrm{coll}$ of the optical microscope is estimated at $\eta_\textrm{coll}=S_\textrm{sub}\times T_\textrm{obj}\times \Gamma_\textrm{APD}=0.11$ where $S_\textrm{sub}=0.41$ is the portion of the total solid angle collected by the high NA objective, $T_\textrm{obj}=0.7$ is the transmission of the objective and $\Gamma_\textrm{APD}=0.4$ is the quantum efficiency of the detector.

\begin{acknowledgement}

This work has been partially funded by the French Agence Nationale de la Recherche (ANR-20-CE24-0001 DALHAI and ISITE-BFC ANR-15-IDEX-0003), the EIPHI Graduate School (ANR-17-EURE-0002). ETH acknowledges support of the Werner Siemens-Stiftung (WSS). Device characterization was performed at the technological platforms SMARTLIGHT and ARCEN Carnot with the support of the French Agence Nationale de la Recherche under program Investment for the Future (ANR-21-ESRE-0040), the Région de Bourgogne Franche-Comté, the European Regional Development Fund (FEDER-FSE Bourgogne Franche-Comté 2021/2027), the CNRS and the French RENATECH+ network. The authors thank Mark S. Sonders (Columbia University, NY) for his critical reading of the manuscript.

\end{acknowledgement}


\end{bibunit}
\section{Supplementary sections}
\begin{bibunit}

\subsection{Experimental details}

\begin{figure}
    \centering
    \includegraphics[width=0.7\linewidth]{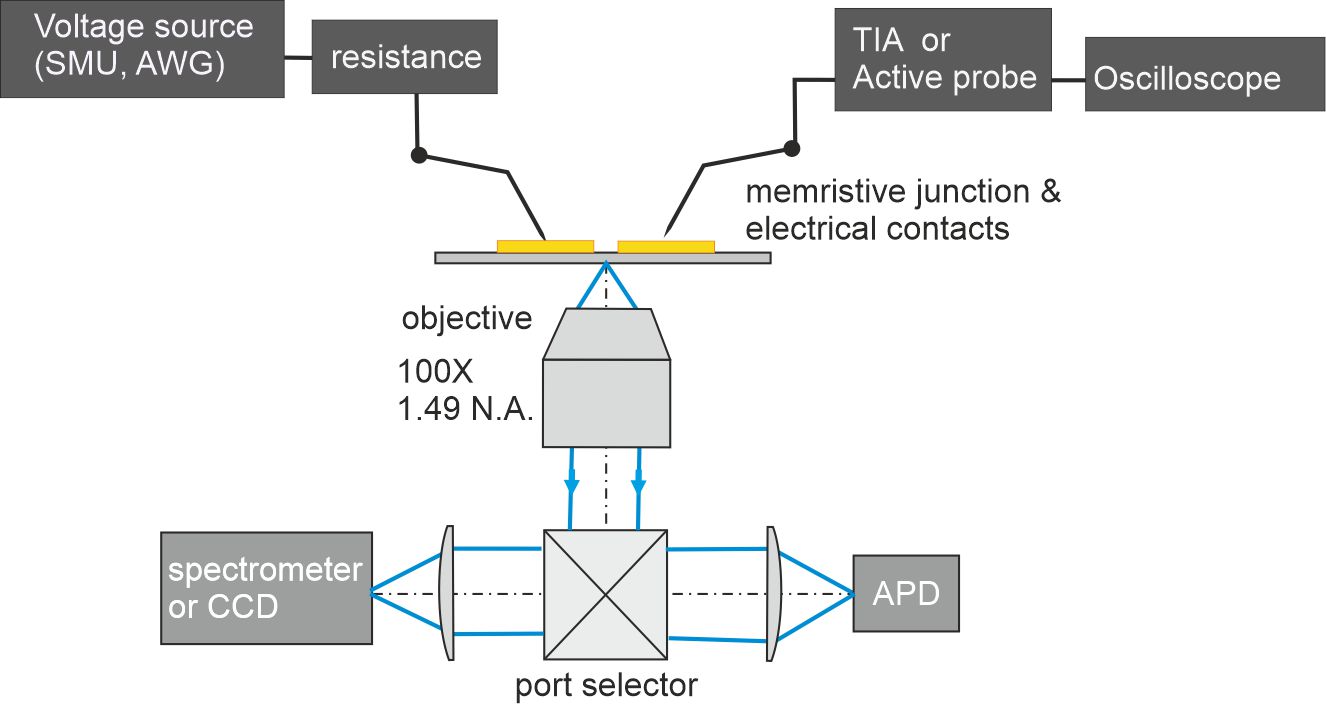}
    \caption{Details of the experimental configuration. In the pulsed measurements, the resistance is integrated with the active probe and no external resistance is present.}
    \label{fig:setup}
    \end{figure}

The memristors are placed on an inverted microscope (Nikon Eclispe) equipped with high numerical aperture objective (Nikon, 100$\times$, NA=1.49). The devices are operated with a controllable voltage source (Keithley 4200). Depending on the targeted time resolution of the experiment, the current is detected by either a transimpedance amplifier (Femto, GmbH) or an active high speed probe (GGB Industries, Inc.) with an input impedance of $R_\textrm{{in}}=1.25$~M$\Omega$. The light emitted by the device and collected by the objective is either imaged on a CCD camera (Andor Newton) and its spectral content analyzed by a spectrometer (Andor Shamrock), or sent to a single photon avalanche photodiode counter (Excelitas). Figure~\ref{fig:setup} shows a sketch of the experiment.

\paragraph{Voltage on DUT:}The current is measured by an active probe with an input impedance $R_\textrm{{in}}=1.25$~M$\Omega$ acting as a voltage divider. Hence, $V_\textrm{{DUT}}$ depends on the current flowing and can be estimated from the relation $V_\textrm{{b}}-R_\textrm{{in}}I$. For all the measurements shown in the main text, the data are acquired for single 10~ms pulses of varied amplitude $V_\textrm{{b}}>V_\textrm{TH}$. An example of the time traces of the recorded signals is illustrated in Fig.~3 of the main text.
\paragraph{Time resolution of the measurements:}
The data are acquired with a 10 ns bin time, this allows to analogically detect the TTL Pulse of the APD. To reduce noise the data are smoothed by a moving average with a window size of 5 $\mu$s.

\subsection{Electroforming phase}

\begin{figure}
    \centering
    \includegraphics[width=\linewidth]{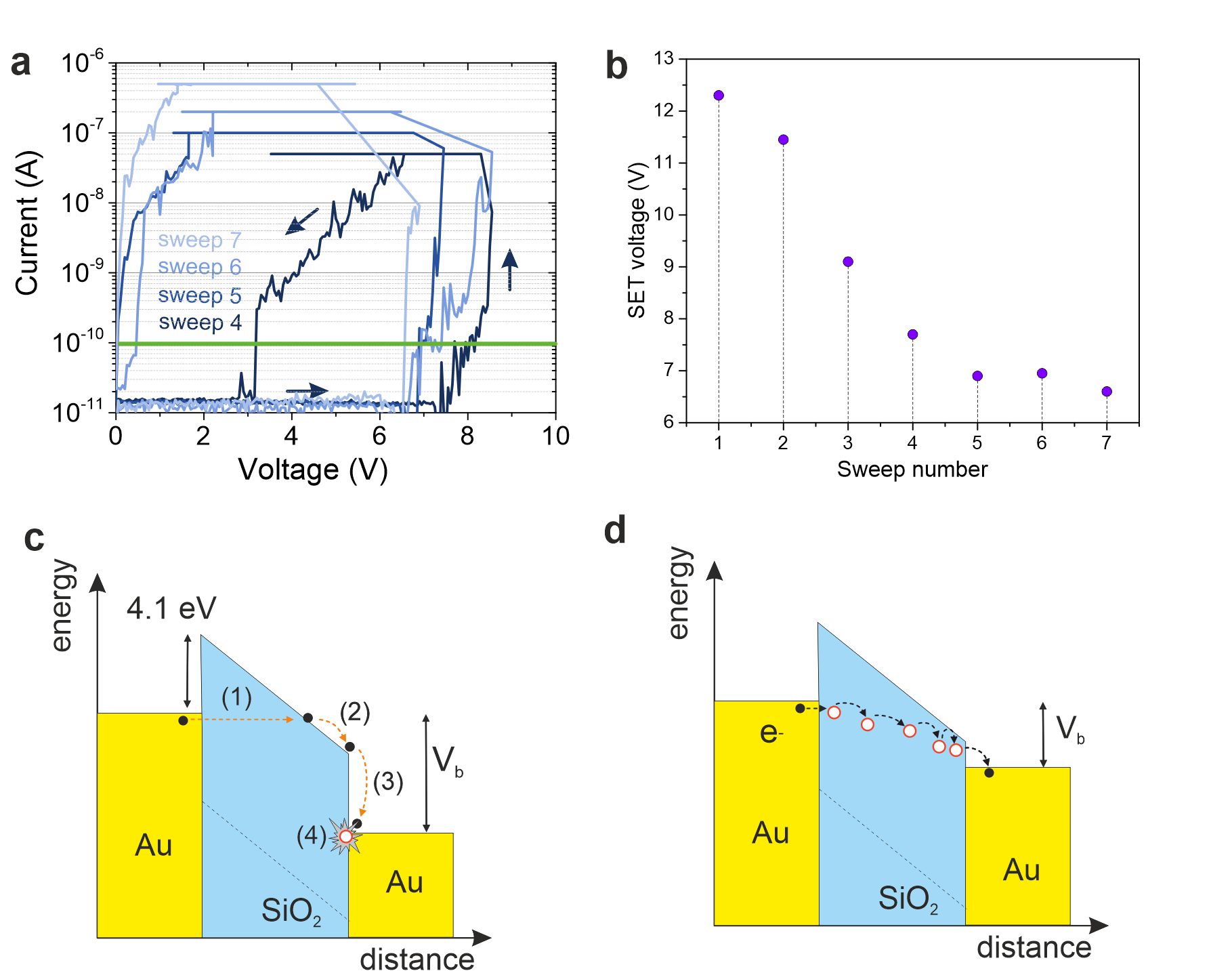}
    \caption{Electroforming of the device. (\textbf{a}) is the current-voltage characteristics taken after the initial sweep taken at a constant $I_\textrm{cc}$.  Here $I_\textrm{cc}$ is increased at each sweep to increase the defect density. \textbf{b}) Evolution of the voltage threshold $V_\textrm{TH}$ for which the device undergoes a resistive switch. $V_\textrm{TH}$ is arbitrarily defined as the voltage at which a current of $I=100$~pA is flowing through the device (green lines in (a)). (\textbf{c}) Sketch of the process at play during early stage of electroforming. Electrons are tunneling to the conduction band of \ce{SiO2} (step 1). The excess energy released (step 2 and step 3) contributes to the creation of defects and electron traps in the matrix (step 4). (\textbf{d}). These defect states promote electron conduction in the dielectric through a hopping-assisted tunneling process.}
    \label{fig:VTH}
    \end{figure}

During the starting sweep discussed in Fig.~1 (b)) of the main text, no measurable current is flowing in the device until the bias $V_\textrm{{b}}$ reaches a threshold voltage $V_\textrm{{TH}}=12.3$~V. $V_\textrm{{TH}}$ marks the onset of resistive switching and is defined here when the current $I$ exceeds 100~pA (approximately 5$\times$ the noise floor, green lines). $I$ then increases dramatically to reach the compliance current ($I_\textrm{cc}=10$~nA) set to protect the device from an irreversible destruction. $I=I_\textrm{cc}$ is maintained during the backward sweep (see arrow) until $V_\textrm{{b}}\sim$ 4~V. The current then decreases to its noise level as the device relaxes back to a low conductance state.  As shown in Fig.~\ref{fig:VTH}(a), $V_\textrm{TH}$ takes lower values at each consecutive sweeps and is indicative of increased defect density present in the oxide~\cite{Sune12}. Figure~\ref{fig:VTH}(d) illustrates the effect of a repeated electrical stress where a systematic decrease of $V_\textrm{{TH}}$ with chronologically ordered sweeping runs is observed. Due to a lower $V_\textrm{{TH}}$, higher compliance currents can be sustained by the device as depicted in the sweeps in Fig.~\ref{fig:VTH}(a).  This electrical behavior is typically explained within the oxide breakdown framework~\cite{Wolters1996}. 

During the voltage actuation, electrons from the source electrode are injected 
into the conduction band of \ce{SiO2} by Fowler-Nordheim tunneling as sketched in Fig.~\ref{fig:VTH}(c), step (1). These electrons then drift towards the counter electrodes (step (2)) where they enter the metal with a significant potential energy difference (4.1~eV for a flat electrode surface \cite{Quattropani1998}, step (3)). Some of this excess energy contributes to the generation of various point defects near the dielectric interface (step (4))~\cite{Lombardo_05}. Additionally, the electric field also causes a series of electrochemical reactions both at the gold electrodes and the dielectric layer, favoring migration of Au ions and oxygen vacancies into the switching matrix \cite{Peng2012,Fowler2015,Inoue2017,Lubben2019,Cheng2022}.Once a sufficient defect density is reached, a current flows in the dielectric medium by electrons hopping between localized states~\cite{Mehonic_2012,Menzel19} (Fig.~\ref{fig:VTH}(d))  and the device switches its conductance state. 

\subsection{Potentiation and depression behaviors}
\begin{figure}
    \centering
    \includegraphics[width=\linewidth]{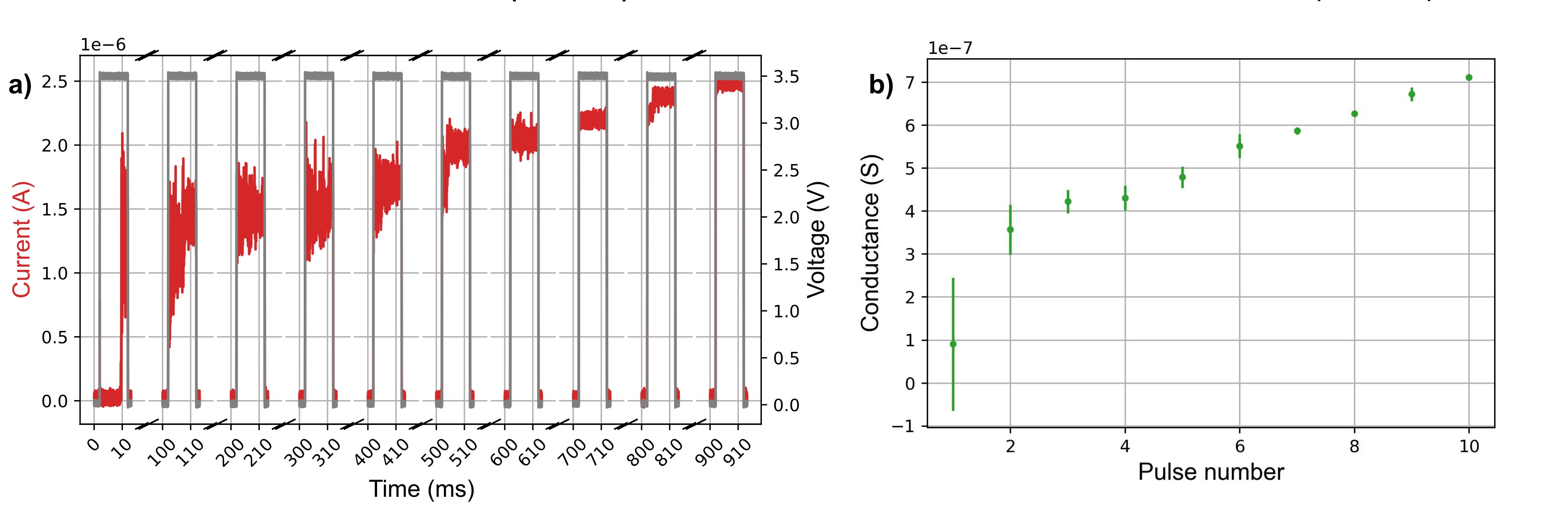}
    \caption{(a) Evolution of the current (red) over 10 consecutive voltage pulses of 3.5 V amplitude and 10 ms duration and 100 ms repetition period. For a greater clarity, the current is traced only when pulses are applied. b) Inferred mean conductance (green) measured for each pulse versus pulse number. The error bars are one standard deviation.}
    \label{fig:Potentation}
    \end{figure}

The conductance value of the volatile state results from a subtle balance between the forming of current pathways and their thermodynamic relaxation. It is influenced by past electrical history and the voltage-induced filament driving force.  This conductance dynamics is revealed in potentiation-depression characteristics of the memristive device, analogous to what is observed in biological neuronal cells \cite{jimbo1999simultaneous}.
The potentiation behavior is illustrated in Figure \ref{fig:Potentation} (a) where a series of 10 consecutive voltage pulses (3.5 V amplitude, 10 ms duration and 100 ms repetition period) is applied. The current gradually increases at each pulses. Figure \ref{fig:Potentation} (b) shows the mean conductance (averaged from the current time trace during the pulses) as a function of pulse number. This clear rise of the conductance with repeated pulses reproduces the potentiation of a synapse subject to an increased activity~\cite{jimbo1999simultaneous}.\\ 

\begin{figure}
    \centering
    \includegraphics[width=\linewidth]{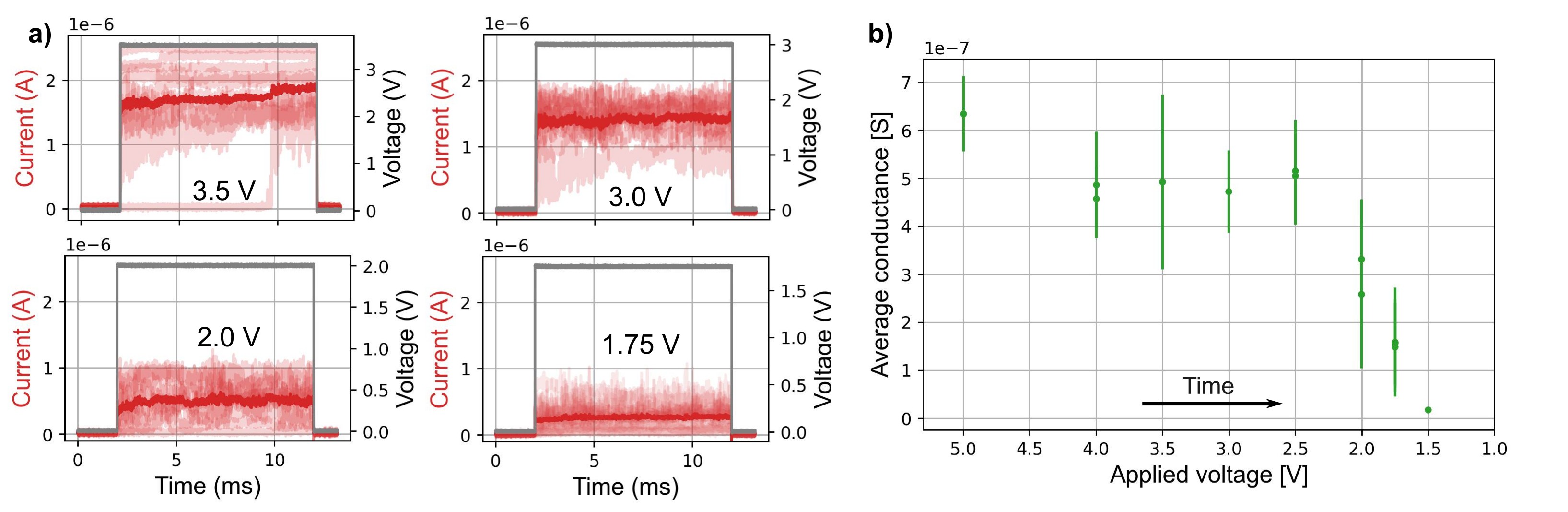}
    \caption{a) Sequence of varying pulse amplitudes, all with 10 ms duration and their associated currents concatenated to a single pulse. The current measured in each pulse is shown in the light red curves, the overall average is shown by the dark red current. b) Average conductance versus the applied voltage. The error bars indicate the variance of the current with respect to the mean. The device was first subjected to high voltages which were then decreased in the next steps. }
    \label{fig:Depression}
    \end{figure}
Synaptic weakening occurring in neuronal systems can also be retrieved from the device volatile behavior. Figure~\ref{fig:Depression}(a) shows the current time traces measured during four pulse sequences with decreasing pulse amplitudes. Each sequence has been concatenated to a single pulse. The dark red trace is the averaged time trace of the entire sequence. In Figure~\ref{fig:Depression}(b), we extract the mean conductance for each sequence and plot it as a function of the pulse amplitude. We clearly observe over time a decrease of the level of the conductance with lowering biases.

\subsection{Numerical simulation of the response of the optical antenna}
\begin{figure}
    \centering
    \includegraphics[width=1\linewidth]{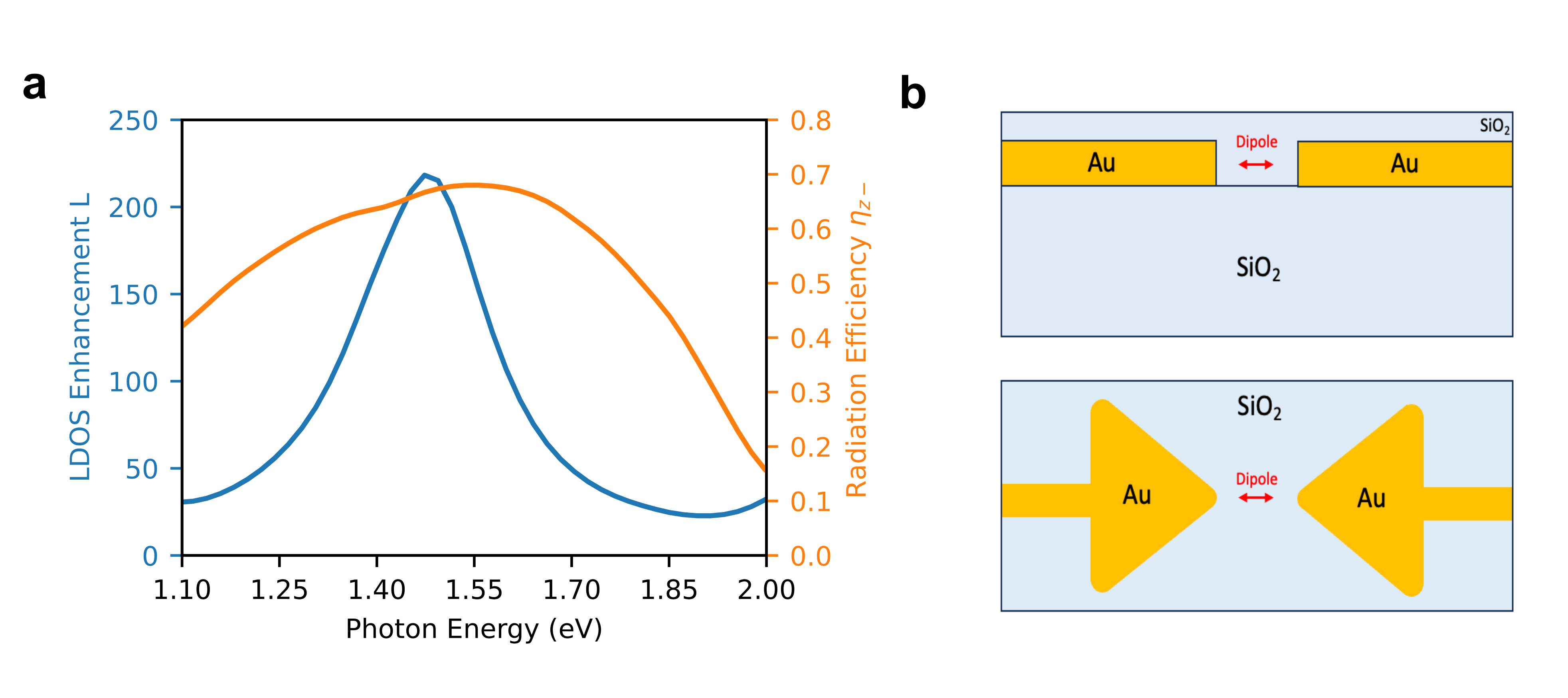}
    \caption{(\textbf{a}) FDTD simulation of the LDOS enhancement L and the radiation efficiency $\eta_{z-}$ perfomed in Ansys Lumerical. The introduction of the antenna results in an increase of the LDOS enhancement up to $\sim218$ and a radiation efficiency $\eta_{z-}$ up to $\sim0.7$. This causes a significant increase of the total quantum efficiency of the system up to a factor $\sim 291$. (\textbf{b}) Simulation setup. A radiating dipole is placed at the midpoint between the electrodes. The electrodes are structured as optical bowtie antennas to enhance the quantum efficiency of the photon emission.}
    \label{fig:Simulation}
\end{figure}

To optimize the quantum efficiency of the memristor light source, the gold electrodes were structured as optical bowtie antennas with a resonance around the photon energy of the observed emission. The bowtie antennas were designed with the help of finite difference in time domain (FDTD) simulations in Ansys Lumerical. In this section we discuss the influence of the electrode structuring on the emission characteristics. \newline
In the approximation of the emissive trap states in the \ce{SiO2} as two level systems (TLS), the total quantum efficiency ($QE$) is given by the following formula:
$$QE=QE_i \cdot L\cdot \eta $$
where $QE_i$ is the internal quantum efficiency, L is the local density of optical states enhancement $L$ (LDOS enhancement) and $\eta$ is the radiation efficiency. $QE_i$ is an intrinsic quantity of the emitter and describes the relative fraction of the radiative decay rate $\gamma_r$ compared to the non-radiative decay rate $\gamma_{nr}$ in the TLS given by $QE_i=\frac{\gamma_{r}}{\gamma_r + \gamma_{nr}}$. The LDOS enhancement $L$ denotes the multiplication factor by which the radiative decay rate $\gamma_{rad}$ is increased by the local environment compared to the case in vacuum. Finally, $\eta$ describes what fraction of the emitted photons are radiated outwards versus absorbed in the electrodes. As the measurement setup only captures light that is emitted towards the z- direction, we can introduce a radiation efficiency $\eta_{z-}$ that is defined as the fraction of the light that is emitted towards the z- direction. Accordingly, we can also define a new quantum efficiency:
$$QE_{z-}=QE_i \cdot L\cdot \eta_{z-} $$
Figure \ref{fig:Simulation} shows the simulated LDOS enhancement and the radiation efficiency for the gold electrodes structured as bowtie antennas. As depicted in the inset the simulations were performed with a radiating dipole polarized along the antenna axis that is placed at the midpoint between the two gold electrodes to represent the average position of an electroluminescent defect. As evident from the simulations, structuring the electrodes creates an LDOS enhancement with a maximum value of around $\sim218$ at resonance as well as a high radiation efficiency $\eta_{z-}$ over the whole energy range of the measured photon emission of 1.2-1.7eV. Compared to the photoemission of a radiating defect center in vacuum the vicinity of the gold bowtie antennas results in an increase of the quantum efficiency $QE_{z-}$ of a factor $\sim59$ up to $\sim291$ in the energy range of the measured electroluminescence.

\subsection{Emission spectra from another device}
\begin{figure}
    \centering
    \includegraphics[width=0.5\linewidth]{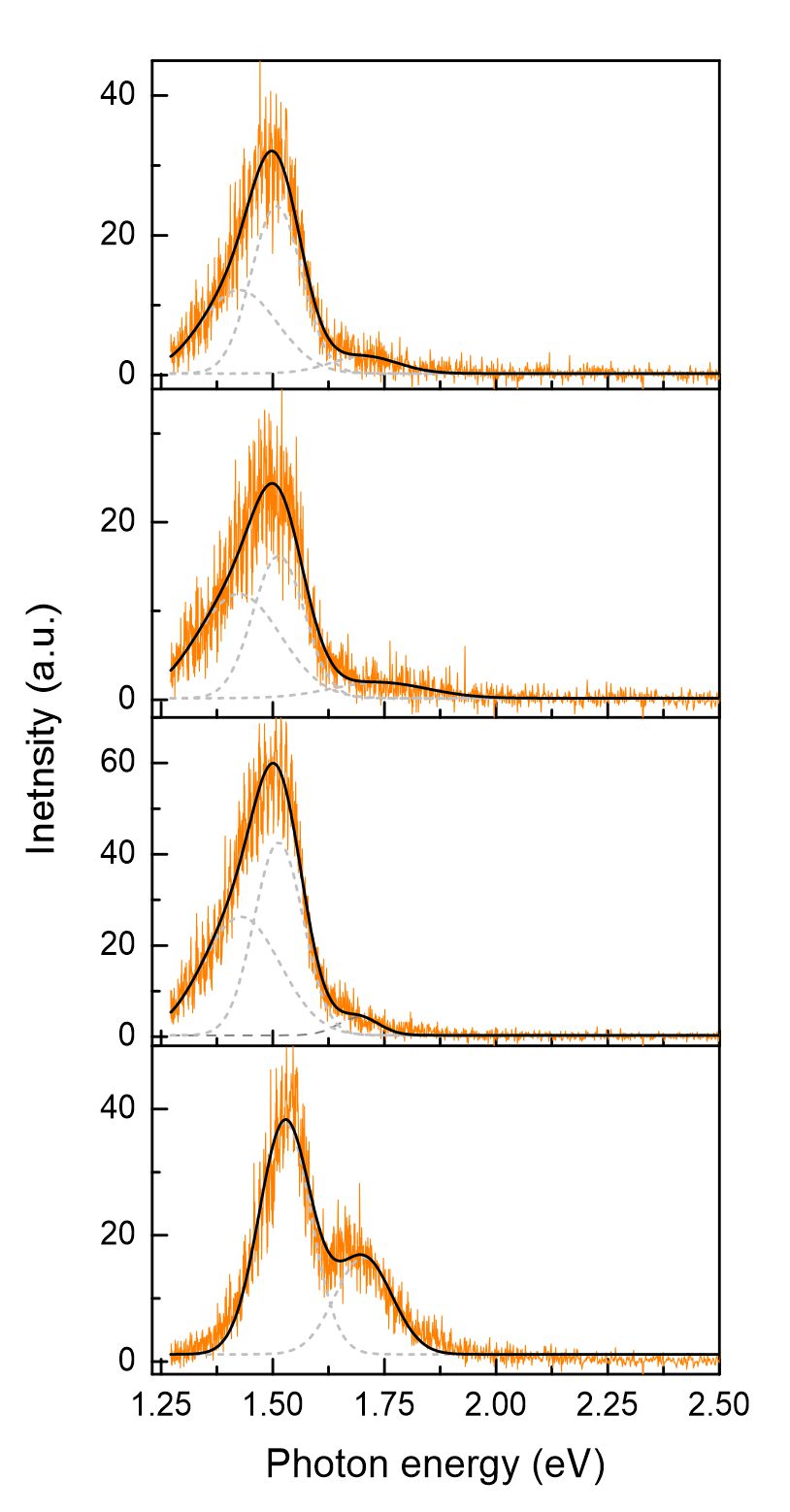}
    \caption{Emission spectra. Series of consecutive spectra acquired from a memristor operated under a constant bias stress $V_\textrm{b}=13$~V. The device is exposed to ambient air and is electrically protected by a 100 M$\Omega$ resistance. Spectra are acquired with a 5~s integration time. The spectral shape can be decomposed into three peaks (dashed curves) with changing weights between each acquisition. The black lines are resulting fits of the spectral decomposition.}
    \label{fig:EL_air}
    \end{figure}

In this section we present emission spectra released by another device featuring similar emission coverage. The overall contact geometry of the memristor is similar to the one discussed in the main text. The electroformation phase followed the same general guidelines with ramping stress voltages and observation of a reducing switching threshold voltage $V_\textrm{TH}$ with stress cycles. In this particular example, the memristor is exposed to air, i.e., there is no encapsulating \ce{SiO2} layer. Hence, the conductive path is most likely formed in the glass substrate. We introduced a 100~M$\Omega$ protection resistance in series with the device. Because the resistance acts as a voltage divider, a higher bias is thus required to operate the device and observe light emission. Figure~\ref{fig:EL_air} shows a series of consecutive spectra released by the memristor when $V_\textrm{b}=13$~V. Each spectrum is integrated for 5~s. The emission here covers a similar spectral band than the device discussed in the main text. We performed a spectral decomposition to analyse the emission peaks. The black lines are the results of the fits and the grey dashed curves are the modes used. All the spectra can be decomposed in a main peak at 1.51~eV flanked by two satellites at 1.42~eV and 1.7~eV whose weight changes between spectra. Each contributing peak can be associated with a specific size of the recombining Si centers~\cite{Liu:16}. The main peak is thus likely emitted by a population of nanocrystals with approximately 4 nm diameter, while the shoulder at high energy stems from a slightly lower Si clusters (3 nm). The low energy peak can be attributed to 5~nm centers.   

\subsection{Telegraph noise}
\begin{figure}
    \centering
    \includegraphics[width=0.7\linewidth]{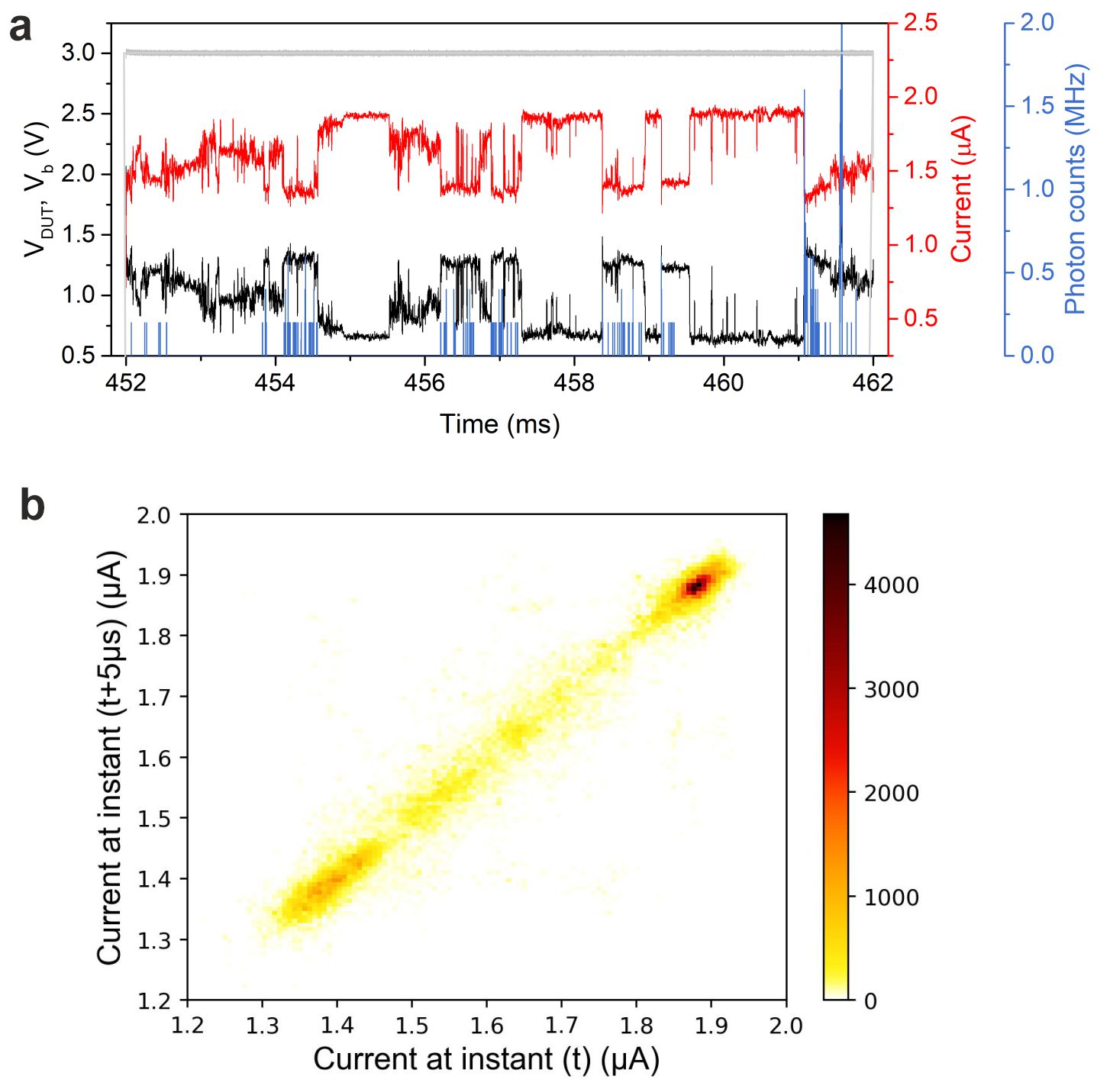}
    \caption{Telegraph noise. (\textbf{a}) Time traces showing the applied bias $V_\textrm{{b}}$ (grey), the current (red), the photon counts (blue) and the voltage $V_\textrm{{DUT}}$ (black) dropped at the device for a pulse featuring well-defined telegraph noise. (\textbf{b}) Current time lag plot extracted from the current evolution featuring two constellations.}
    \label{fig:SM_pulse9}
    \end{figure}

Some of the voltage pulses applied to the memristor features telegraph noise demonstrating that transport in the dielectric matrix is ruled by a very few traps. This is exemplified in Fig.~\ref{fig:SM_pulse9}(a) showing the evolution of the different signals over a 10 ms voltage pulse. Clearly, the current is evolving towards two-well defined conductive states. The current time lag plot displayed in Fig.~\ref{fig:SM_pulse9}(b) confirms the presence of two marked constellations at $I=1.4~\mu$A and $I=1.9~\mu$A which are the signature of a single active charge trap~\cite{Ohata:17}. The photon time trace is perfectly in line with the observation because the device releases photons only when electrons are trapped in the defect.

\newpage
\subsection{Current correlation}
\begin{figure}
    \centering
    \includegraphics[width=0.75\linewidth]{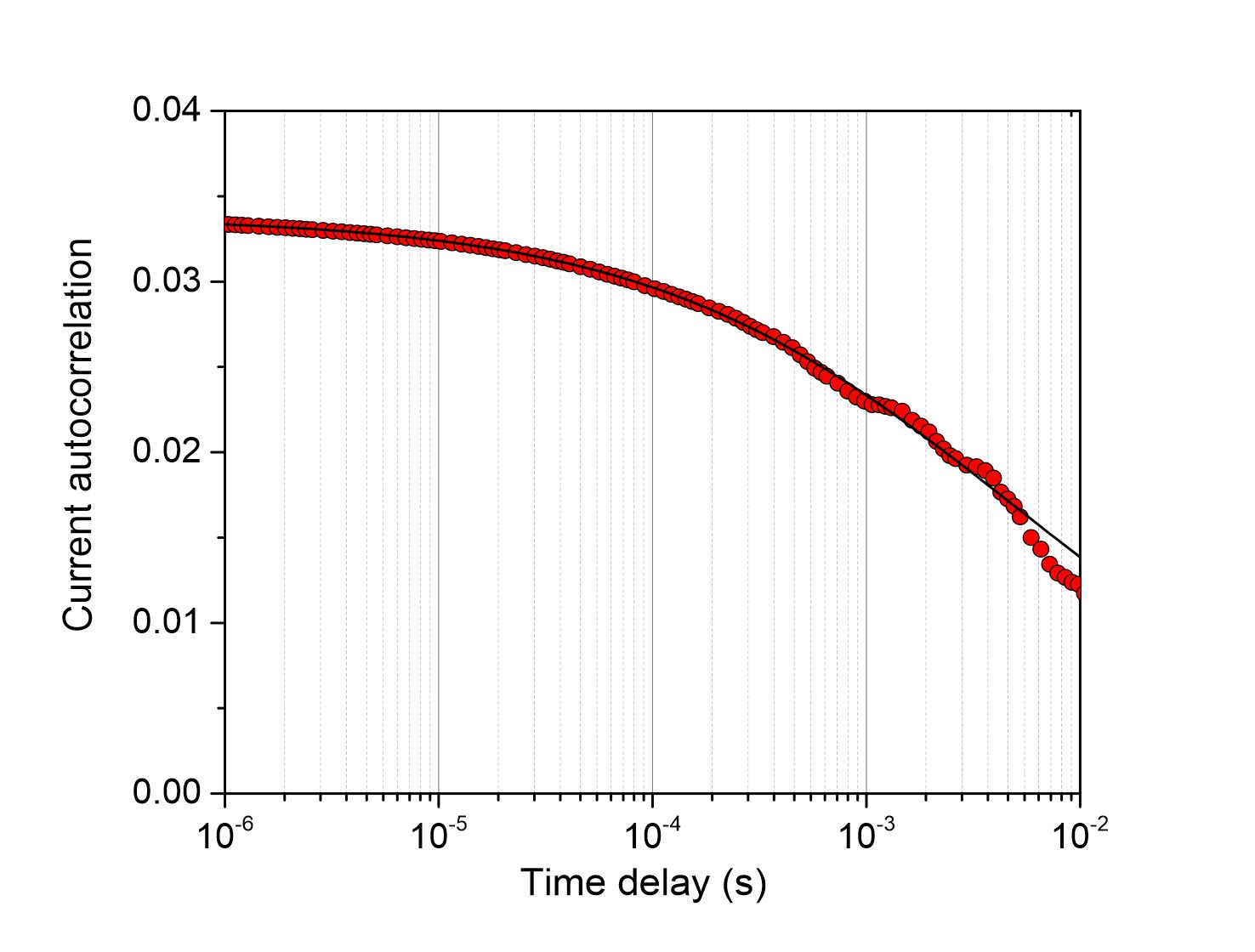}
    \caption{Second order autocorrelation $g^{(2)}(\tau)$ of the current trace. The black line is a fit with an anomalous diffusion model.}
    \label{fig:autoco_current}
    \end{figure}

Figure~\ref{fig:autoco_current} is the autocorrelation of the current time trace shown in the main body. Because the fluctuations of both photon emission and current transport are ruled by electron trapping dynamics, we tentatively used the anomalous diffusion model to extract quantitative parameters. The black line is the fit with parameters provided in Table~\ref{tab:fits_current}.

The active time $\tau_a$ is very similar to the one deduced from the photon autocorrelation discussed in the main body ($\tau_a$=0.42~ms vs 0.33 ms for the photon autocorrelation). The fit of the current autocorrelation renders a slightly higher sub-diffusive exponent, which is understood from the fact that the distribution of hopping steps responsible for the photon variation and current fluctuation is not the same. 
The fit also provides a larger number of defect centers and charge traps participating in the current conduction. This is expected as the luminescent defects contribute also to current conduction but most of the defects are non-luminescent. However, in the absence of a formal transport model equivalent to fluorescent correlation spectroscopy, the meaning of $N$ is disputable.  

\begin{table}
    \centering
    \begin{tabular}{c|c|c|c}
         \(N\) &  \(\tau_a\)  & \(\beta\)  \\
        \hline
         603 & 0.33 ms & 0.61 \\
    \end{tabular}
    \caption{The anomalous diffusion fit parameters deduced from Fig.~\ref{fig:autoco_current}. The numbers in brackets indicate the fit uncertainty in the last digits. }
    \label{tab:fits_current}
\end{table}

\subsection{Photon autocorrelation with a constant voltage bias}

\begin{figure}
    \centering
    \includegraphics[width=0.75\linewidth]{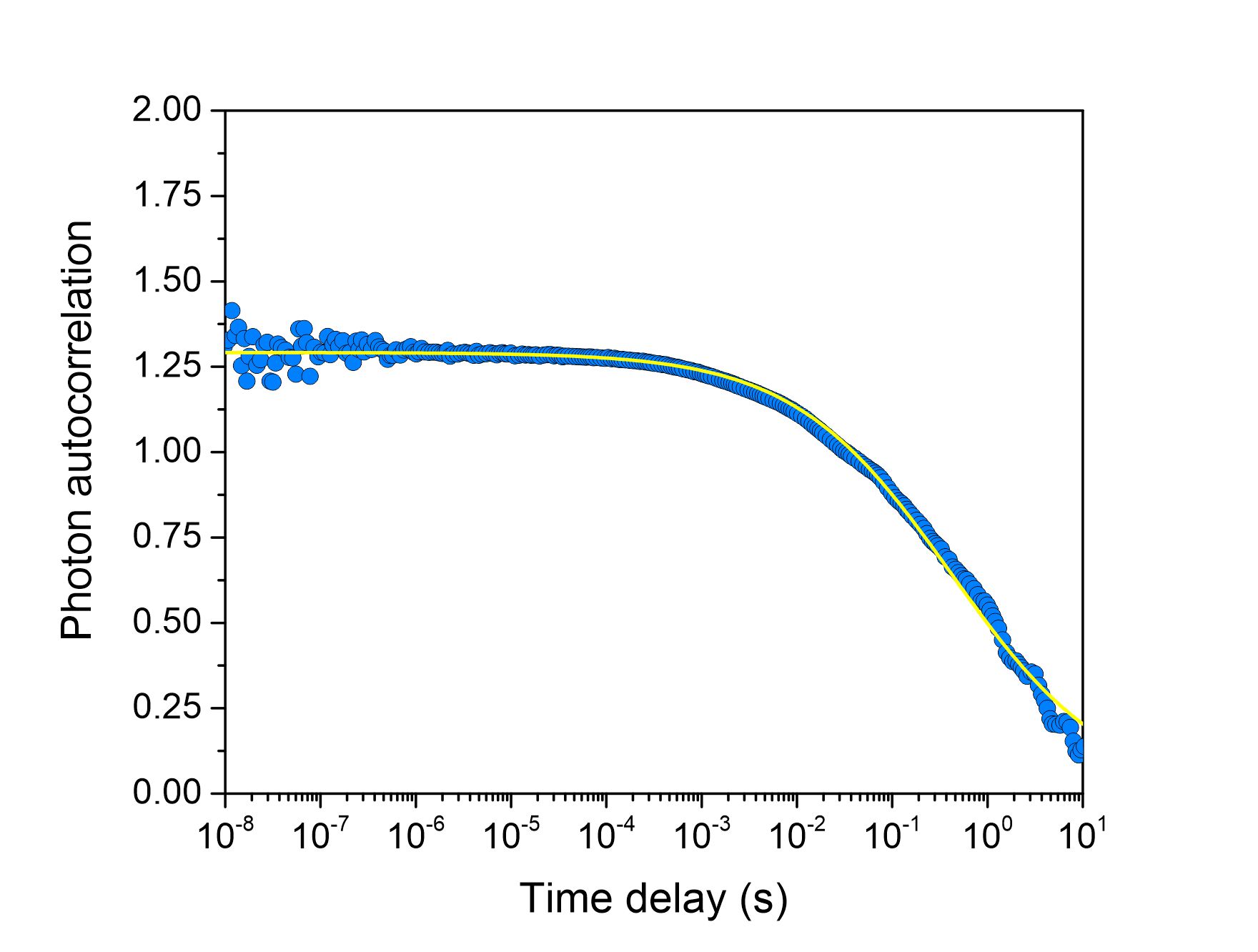}
    \caption{Second order autocorrelation $g^{(2)}(\tau)$ of the photon rate for a junction operated in air under a constant bias. The yellow line is a fit with the anomalous diffusion model.}
    \label{fig:autoco_CW}
    \end{figure}

In this section, we show that the number of emitting centers produced in the memristor stay sparse even under a constant bias actuation. In this experiment, the junction is operated in air at $V_\textrm{b}=4$~V for about 400~s and is protected by a 11~M$\Omega$ protection resistance. For this measurement, the autocorrelation of the photon counts was measured using a Handburry Brown and Twiss interferometer with the help of second APD and a correlator.  Figure~\ref{fig:autoco_CW} shows the correlation curve of the photon rate calculated from the time traces recorded by the APDs. The yellow line is a fit using the anomalous diffusion model discussed in the main text with parameters given in Table~\ref{tab:fits_CW}. Light emitted by this device also stems from less that one recombining center $N$ in average. This statement is consistent with previous analyses discussing fluctuating spectral contributions and photon intermittency.

\begin{table}
    \centering
    \begin{tabular}{c|c|c|c}
         \(N\) &  \(\tau_a\)  & \(\beta\)  \\
        \hline
         0.77(1) & 410(50) ms & 0.52(4) \\
    \end{tabular}
    \caption{The anomalous diffusion fit parameters deduced from Fig.~\ref{fig:autoco_CW}. The numbers in brackets indicate the fit uncertainty in the last digits. }
    \label{tab:fits_CW}
\end{table}

We note a much longer $\tau_a$ than the one inferred for the memristor covered by \ce{SiO2} discussed in the main text. This can partially explained by the nature of the excitation. Whereas the junction in air under constant bias is all the time in a low resistance state and electrically stressed, the memristor embedded in \ce{SiO2} is excited by pulses (10 ms duration) and thus switches repeatedly between resistive states. The pulsed excitation thus promotes a faster diffusion dynamics of the conduction path compared to a constant excitation.


\end{bibunit}

\end{document}